\documentclass[prc,
 amsmath,amssymb,
 aps,superscriptaddress
]{revtex4-2}

\usepackage{amsmath}
\usepackage{dcolumn}
\usepackage{graphicx}
\usepackage[utf8]{inputenc}
\usepackage[caption=false]{subfig}
\usepackage{hyperref}
\usepackage{ulem}

\hypersetup{
colorlinks=true
}

\graphicspath{{images/}}

\begin{document}

\preprint{}

\title{Initial Baryon Stopping and Angular Momentum in Heavy-Ion Collisions}

\author{Alex Akridge}
\email{zakridge@iu.edu}
\affiliation{ Physics Department and Center for Exploration of Energy and Matter,
Indiana University, 2401 N Milo B. Sampson Lane, Bloomington, IN 47408, USA.} 

\author{Daniel Gallimore}
\affiliation{ Physics Department and Center for Exploration of Energy and Matter,
Indiana University, 2401 N Milo B. Sampson Lane, Bloomington, IN 47408, USA.}

\author{Hector Morales}
\affiliation{Hendrix College, 1600 Washington Ave, Conway, AR 72032, USA.}
\affiliation{ Physics Department and Center for Exploration of Energy and Matter,
Indiana University, 2401 N Milo B. Sampson Lane, Bloomington, IN 47408, USA.} 

\author{Jinfeng Liao}
\email{liaoji@indiana.edu}
\affiliation{ Physics Department and Center for Exploration of Energy and Matter,
Indiana University, 2401 N Milo B. Sampson Lane, Bloomington, IN 47408, USA.}

\date{\today}

\begin{abstract}
Noncentral heavy-ion collisions create fireballs with large initial orbital angular momentum that is expected to induce strong vorticity in the hot bulk fluid and generate global spin polarization of the produced particles. As the collision beam energy $\sqrt{s_{\rm NN}}$ decreases to approach the two-nucleon-mass threshold, this initial angular momentum approaches zero. One may thus expect that the observed global spin polarization should reach a maximum and then drop to zero as increased stopping competes with decreased initial momentum. Recent experimental measurements, however, appear to show a continual rise of hyperon polarization even down to $\sqrt{s_{\rm NN}} =$ 2.42 GeV, suggesting a peak very near threshold which is difficult to interpret and calls for a better understanding of angular momentum initial conditions, especially at low energy. Here, we develop a new Glauber-based initial state model (``Glauber+'') to investigate the initial distribution of angular momentum with respect to rapidity as well as the dependence of this distribution on initial baryon stopping across a wide range of collisional beam energy. We estimate that the angular momentum per produced final charged particle at mid-rapidity peaks around 5 GeV, which may present a potential challenge to an interpretation of the spin polarization measurements near threshold as being a consequence of the initial angular momentum of the colliding system.
\end{abstract}

\maketitle

\section{Introduction} \label{sec:intro}
In a non-central heavy-ion collision at high energy, a large amount of initial orbital angular momentum (OAM) is present in the colliding system, part of which is carried by the created fireball at mid-rapidity. It is natural to ask what happens during the subsequent transport of this angular momentum and, in particular, whether there could be some experimentally observable consequences. 
It was proposed in  {Ref.} \cite{Liang:2004ph} that a global hyperon polarization could be induced along the direction of the initial OAM (perpendicular to the reaction plane spanned by the beam direction and impact parameter) as an observable. 
It was later shown that a nonzero OAM leads to the formation of  {a} nontrivial thermal vorticity field in a fluid dynamics description which then further causes spin polarization of underlying particles~\cite{Becattini:2007sr,Becattini:2013fla,Betz:2007kg}. Experimentally, early STAR measurements~\cite{STAR:2007ccu} at 62.4 and 200 GeV for $\Lambda$ and $\bar{\Lambda}$ baryons in Au+Au collisions suggested a small or zero value for this polarization. In 2017, STAR reported a global polarization for $\Lambda/\bar{\Lambda}$ at a few percent level~\cite{STAR:2017ckg} for $\sqrt{s_{\rm NN}} < $ 40 GeV, indicating an extremely large vorticity $\sim 10^{22} \textrm{ s}^{-1}$ in the QGP. This discovery further motivated many subsequent measurements on both global and local spin polarizations as well as spin alignment of vector mesons~\cite{STAR:2018gyt,STAR:2020xbm,STAR:2021beb,STAR:2023eck,STAR:2023nvo,ALICE:2021pzu,Sarkar:2022uyt,ALICE:2022dyy,STAR:2022fan,Xi:2023iua,Wilks:2024job,Shen:2024rdq,ALICE:2022byg,Bai:2024ubz,Micheletti:2024lqb}. 
Theoretically, a paradigm of ``angular-momentum $\to$ vorticity $\to$ spin polarization'' has been demonstrated by various hydrodynamic and transport models to provide a quantitative description of global spin polarization in heavy ion collisions. For reviews, see e.g.  {Refs.} \cite{Becattini:2020ngo,Becattini:2021wqt,Becattini:2021lfq,Huang:2020dtn,Gao:2020lxh,Niida:2024ntm,Jiang:2021ict,Cao:2021nhi,Karpenko:2021wdm,Becattini:2024uha}.
 
More recently, measurements of global hyperon polarization have been extended by STAR and HADES Collaborations down to a beam energy range of just a few GeV~\cite{STAR:2021beb,HADES:2022enx}. Surprisingly, their results appear to suggest a large polarization signal which continues a  {monotonically} increasing trend even when the beam energy approaches the twice-nucleon-mass threshold (i.e. $\sqrt{s_{\rm NN}} \to 2m_{\rm N}$). Such a trend is difficult to understand if one considers the beam energy dependence of the fireball angular momentum in these collisions. The total initial orbital angular momentum in a collision at impact parameter $b$ can be estimated as $|J_y|  {\sim Abp_z \sim} \frac{1}{2} A b \sqrt{s_{\rm NN}  - (2m_{\rm N})^2}$, where the key factor is the amount of longitudinal momentum per initial nucleon {, $p_z$,} which monotonically decreases with lower and lower beam energy. In particular $J_y$ must approach zero when $\sqrt{s_{\rm NN}} \to 2m_{\rm N}$. 
On the other hand, as $\sqrt{s_{\rm NN}}$ decreases, more matter is stopped at mid-rapidity, so there is a bigger fraction of the total angular momentum carried by the mid-rapidity fireball, where most polarization measurements are made. So the two effects (total angular momentum versus stopping at mid-rapidity) with opposite beam energy dependence will compete against each other, presumably resulting in a non-monotonic $\sqrt{s_{\rm NN}}$ dependence of the fireball angular momentum (which in turn determines the global spin polarization signal) that shall peak at a certain $\sqrt{s_{\rm NN}}$ and then vanish toward the threshold. 
The measurements of STAR (at $\sqrt{s_{\rm NN}} = $ 3 GeV) and HADES (at $\sqrt{s_{\rm NN}} =$ 2.42 GeV), if reliable, would thus suggest a peak of this phenomenon at an energy extremely close to the threshold, which could be challenging to understand quantitatively. Phenomenological model studies in this energy range show mixed results~\cite{Guo:2021udq,Deng:2021miw,Deng:2020ygd,Ivanov:2020udj,Ivanov:2022ble}   
but mostly appear to favor a maximum away from the threshold in the 3-10 GeV range.

Clearly, better theoretical understanding is needed. In particular, it is important to understand how the angular momentum is distributed and deposited to the mid-rapidity fireball right after the initial impacts between the colliding nucleons. In this regard, the nucleon stopping in the longitudinal direction (which could also be quantified via rapidity loss~\cite{BRAHMS:2009wlg}) plays a key role. Thus, the initial rapidity distribution of angular momentum should be strongly correlated with the early baryon stopping which in turn connects with the net baryon numbers of measured final state hadrons in mid-rapidity. This therefore offers an interesting way to relate the two and help constrain the angular momentum initial conditions across a broad range of collisional beam energies.

In this paper we investigate the effect of baryon stopping on angular momentum in the initial state of Au-Au collisions. We extend the Glauber model to a ``Glauber+'' model with information just after the initial impacts for the estimates of the initial net baryon as well as angular momentum densities in the transverse plane. The stopping effect is implemented phenomenologically based on constraints from generic conservation laws, with its strength estimated using two empirical parameters that we calibrate with experimental net-proton yields. Our model predicts the rapidity density of baryon number and angular momentum in the initial state. We scan $\sqrt{s_{\rm NN}}$ values in the range $2\sim 200$ GeV and a broad range of impact parameters.

The rest of this paper is organized as follows. In Section\autoref{sec:model}, we define and discuss our model for the deposition of baryon number and angular momentum in the initial state, along with details on the implementation of numerical calculations. Results are presented in Section\autoref{sec:results}, followed by a conclusion in Section\autoref{sec:conclusion}. An Appendix\autoref{sec:baryonfactor} is also included for details on the estimation of net baryon numbers from experimentally measured net proton numbers. 

\section{From Glauber to Glauber+ Model} \label{sec:model}

Let us consider two colliding nuclei, labeled $A$ and $B$, with $A$ ($B$) traveling in the $+z$ ($-z$) direction. While the discussions below can be applied to any colliding systems, we will focus on \textsuperscript{197}Au-\textsuperscript{197}Au collisions, which are measured in the RHIC Beam Energy Scan program~\cite{Bzdak:2019pkr}. Each nucleon (taken to have mass $m_{\rm N} = 939$ MeV) has energy $\sqrt{s_{\rm NN}}/ 2 $. Note that in this paper, we will always use capital $Y$ to denote rapidity values, to avoid confusion with the spatial coordinate $y$. Thus, the nuclei will initially have rapidity $\pm [Y_{\rm beam} = \cosh^{-1}(\sqrt{s_{\rm NN}} / 2m_{\rm N})]$. Note also that we take angular momentum to be in units of $\hbar$. The centers of the nuclei are displaced by an impact parameter $b$, with the direction of displacement taken as the $x$-axis. The $A$ and $B$ nuclei are centered at $(-b / 2, 0)$ and $(b / 2, 0)$ respectively within the transverse plane.

The optical Glauber model is used to calculate various quantities in the initial conditions~\cite{Miller:2007ri}. The nuclear density is modeled with a Woods-Saxon function~\cite{DeVries:1987atn} normalized by the requirement that
\begin{equation}
    \int d^3\vec{r} \rho(|\vec{r}|) = A = 197
\end{equation}
and the 3-dimensional density $\rho$ is integrated over the $z$-axis to obtain the 2-dimensional thickness function $T_{A, B}(x, y)$. The inelastic nucleon-nucleon cross section is calculated from an interpolated formula~\cite{dEnterria:2020dwq}, and the participant densities are given by \footnote{Often the inner bracket is replaced with an exponential since $A$ is large, but it seems there's no reason not to keep the more accurate form.}\cite{Hirano:2005xf}
\begin{equation}
    n_{A, B}(x, y) = T_{A, B}(x, y) \left( 1 - \left( 1 - \frac{\sigma_{NN} T_{B, A}(x, y)}{A} \right)^A \right)
\end{equation}
Thus the integral of $n_A + n_B$ over the $xy$-plane gives the total participant number. For non-central collisions, there is a global orbital angular momentum along the out-of-plane direction, with each nucleon contributing $-xp_z$ to $J_y$. So one can define an initial angular momentum density on the transverse plane as:  
\begin{equation}
    \frac{d^2J_y}{dxdy} = -x \sqrt{\frac{s_{\rm NN} - (2 m_{\rm N})^2}{4}  } \  \left [ n_A(x, y) - n_B(x, y) \right ]
\end{equation} 
where nucleons from the two nuclei contribute oppositely. Integrating the above over the transverse plane gives the total initial angular momentum carried by {\em all participant nucleons}. 

This calculation, however, only reflects the situation just before the collisions. Upon the impact between the two incident nuclei, the momentum values of participant nucleons will be changed immediately after the collisions. In particular, their $p_z$ magnitude would be generally reduced by a varied extent, and thus a {\em rapidity loss} results. The participant nucleons which all travel at the same beam rapidity before the collision will be distributed across a range of different and reduced rapidity values, depending on the details of the collision process. An important consequence is the baryon stopping which transports the net baryon number (which is an exactly conserved quantity) from the initial beam to the mid-rapidity fireball. Obviously, the same effect also leads to the spread out of the initial angular momentum (which also must be conserved) from beam rapidity toward a broad rapidity range. To analyze the non-trivial distributions of baryon number and angular momentum with respect to rapidity, one needs to understand the status of those participant nucleons just after the collision. This requires going beyond the usual Glauber model description of the initial conditions. In the following, we will develop a {\em Glauber+ model}  that describes the distributions of post-collision participant nucleons in the phase space of transverse plane and longitudinal rapidity, which will then allow the calculations of initial rapidity distributions of baryon number and angular momentum.

\label{subsec:elasticity}

\subsection{Collisional elasticity and rapidity loss}

In a typical binary collision of participating nucleons, each nucleon loses a fraction of its initial kinetic energy and longitudinal momentum which turn into transverse motion as well as internal excitations. Such a collision is often highly inelastic. Thus the strength of the stopping effect (due to reduction in longitudinal momentum) is directly related to the elasticity of the collision. To quantify this, we introduce a relativistic version of the coefficient of restitution $e$. It may be noted that non-relativistically an inelastic collision implies loss of kinetic energy, but in special relativity the energy conservation and momentum conservation are non-trivially connected through Lorentz covariance which requires extra caution. One may compare our results here to those of a previous study~\cite{Shen:2020jwv} by Shen and Alzhrani which analyzed the implications of local conservation laws for 3D initial conditions of the energy-momentum tensor.

Let us consider the initial collisions after which each nucleon becomes internally excited as a ``wounded nucleon'' with a mass $m'$ that is greater than the ordinary nucleon mass. Further, each pair of nucleons may deflect each other transversely here, with some momentum transfer $\vec{p}_\perp$. In order to focus on the change of rapidity, we can lump this into an effective mass also, so
\begin{align}
   ( \lambda m_{\rm N} )^2 &\equiv m'^2 + p_\perp^2
\end{align}
where\footnote{For current convenience, $\lambda$ is taken to be a constant common to each binary collision at a given beam energy, and thus reflects an average level of excitation over all possible collision outcomes.} $\lambda m_{\rm N}$ is the new effective mass. With this in mind we can generalize the non-relativistic coefficient of restitution to the current situation. Suppose the initial rapidity values for nucleons from the two incident nuclei are $Y_A, Y_B$ respectively. Right after colliding, the wounded nucleons will have rapidity values $Y_A'$, $Y_B'$ respectively. We define the following coefficient of restitution (COR): 
\begin{align} \label{eq:e}
     e = \frac{Y_B' - Y_A'}{Y_B - Y_A}
\end{align}
Then $e = 1$ corresponds to no interaction, $e = 0$ to a completely inelastic collision, and $e = -1$ to a perfectly elastic collision. We'll let $e$ be a collision-wide free parameter at any given beam energy that will be constrained later. Note also that by definition $e$ is boost-invariant.

Now take an area element $dxdy$ at a specific point $(x, y)$ in the transverse plane, where there are bunches of participating nucleons with total mass $m_{\rm N} n_{A, B}dxdy$ from the $A$ and $B$ nuclei respectively. 
For this area element $dxdy$, we have the corresponding mass elements from each nucleus, $dm_{A, B} = m_{\rm N} n_{A, B}dxdy$. After the collision these become $\lambda dm_{A, B}$. Conservation of $z$-momentum and energy can then be written as
\begin{align}
    \label{eq:mom_conserve}
    dm_A \sinh Y_A + dm_B \sinh Y_B &= \lambda (dm_A \sinh Y_A' + dm_B \sinh Y_B') \\
    \label{eq:en_conserve}
    dm_A \cosh Y_A + dm_B \cosh Y_B &= \lambda (dm_A \cosh Y_A' + dm_B \cosh Y_B')
\end{align}

For simplicity, let us use the global center-of-mass (COM) frame of the incident nuclei A and B, where $Y_A = -Y_B $ and in this case $Y_B' = Y_A' - 2eY_A$ from Eq.~\eqref{eq:e}. Then Eqs.~\eqref{eq:mom_conserve} and~\eqref{eq:en_conserve} can be further simplified as:
\begin{align}
    &(n_A - n_B) \sinh(Y_A) = \notag \\
    &\lambda [(n_A + n_B \cosh(2eY_A)) \sinh(Y_A') - n_B \sinh(2eY_A) \cosh(Y_A')] \\
    &(n_A + n_B) \cosh(Y_A) = \notag \\
    &\lambda [(n_A + n_B \cosh(2eY_A)) \cosh(Y_A') - n_B \sinh(2eY_A) \sinh(Y_A')]
\end{align}
Define quantities $u$ and $K$ by
\begin{align}
    \tanh(u) &= \frac{n_B \sinh(2eY_A)}{n_A + n_B \cosh(2eY_A)} \\
    K &= \lambda \sqrt{n_A^2 + 2n_A n_B \cosh(2eY_A) + n_B^2}
\end{align}
and note that $u$ has the same sign as $e$. Then the conservation laws become
\begin{align}
    (n_A - n_B) \sinh(Y_A) &= K \sinh(Y_A' - u) \label{eq:pconservation} \\
    (n_A + n_B) \cosh(Y_A) &= K \cosh(Y_A' - u) \label{eq:econservation}
\end{align}
 {Dividing Eq.~\eqref{eq:pconservation} by Eq.~\eqref{eq:econservation} and solving for $Y_A'$ yields}
\begin{align} \label{eq:YA}
    Y_A' = &\tanh^{-1} \left( \frac{n_A - n_B}{n_A + n_B} \tanh(Y_A) \right) + u \notag \\
    = &\tanh^{-1} \left( \frac{(n_A/n_B) - 1}{(n_A/n_B) + 1} \tanh(Y_A) \right) + \tanh^{-1} \left( \frac{ \sinh(2eY_A)}{(n_A/n_B) + \cosh(2eY_A)} \right) 
\end{align}
The post-collision rapidity of participant nucleons from nucleus $B$ is then given by
\begin{align} \label{eq:YB}
Y_B' = Y_A' - 2eY_A
\end{align}
Clearly, the respective rapidity loss can be determined from $Y_{A,B}' - Y_{A,B}$, which would be dependent on three factors: the $n_A/n_B$ ratio, the parameter $e$, and the initial beam rapidity. A detailed discussion of two additional useful (albeit technical) points regarding the above equation for rapidity loss calculations has been included in Appendix~\ref{sec:YAB}. 
\vspace{0.2in}

\begin{table}[hbt!]
    \begin{tabular*}{0.75\textwidth}{@{\extracolsep{\fill}}|cccccc|}
    \hline
    $e$ & $\sqrt{s_{\rm NN}} =$ 2 GeV & 3 & 27 & 62.4 & 200 \\
    \hline
    0.8 & 1.02 & 1.17 & 1.95 & 2.31 & 2.92 \\
    \hline
    0.6 & 1.04 & 1.33 & 3.77 & 5.32 & 8.52 \\
    \hline
    0.4 & 1.05 & 1.47 & 7.03 & 11.99 & 24.61 \\
    \hline
    0.2 & 1.06 & 1.56 & 11.65 & 24.19 & 65.25 \\
    \hline
    0 & 1.06 & 1.60 & 14.38 & 33.23 & 106.50 \\
    \hline
    \end{tabular*}
    \caption{Values of the mass ratio $\lambda$ at $w = 0$ or $n_A = n_B$, corresponding to the center of the transverse plane, for different choices of the parameter $e$ and beam energy. The initial rapidity $Y_A$ is set to be the beam rapidity $Y_{\rm beam}$.
    }
    \label{tab:massratiow0}
\end{table}

\begin{table}[hbt!]
    \begin{tabular*}{0.75\textwidth}{@{\extracolsep{\fill}}|cccccc|}
    \hline
    $e$ & $\sqrt{s_{\rm NN}} =$ 2 GeV & 3 GeV & 27 GeV & 62.4 GeV & 200 GeV \\
    \hline
     0.8 & 1.01 & 1.12 & 1.94 & 2.31 & 2.92 \\  
   \hline
    0.6 & 1.02 & 1.23 & 3.67 & 5.27 & 8.50 \\
    \hline
    0.4 & 1.03 & 1.31 & 6.44 & 11.35 & 23.97 \\
   \hline
    0.2 & 1.04 & 1.36 & 9.41 & 19.96 & 55.54 \\
  \hline
    0 & 1.04 & 1.38 & 10.68 & 24.19 & 75.06 \\
    \hline
    \end{tabular*}
    \caption{Values of the mass ratio $\lambda$ at the center of either nucleus with impact parameter $b = 7$ fm, for different choices of the parameter $e$ and beam energy. The initial rapidity $Y_A$ is set to be the beam rapidity $Y_{\rm beam}$.}
    \label{tab:massratiocenterofnucleus}
\end{table}

 {To determine the ratio $\lambda$}, note that an alternative expression for $K$ can be derived from Eqs.~(\ref{eq:pconservation},~\ref{eq:econservation}):
\begin{align}
    K &= \cosh(Y_A) \sqrt{(n_A + n_B)^2 - (n_A - n_B)^2 \tanh^2(Y_A)} \notag \\
    &= \sqrt{n_A^2 + 2n_A n_B \cosh(2Y_A) + n_B^2}
\end{align}
From the definition of $K$, we get
\begin{align}
    \lambda^2 &= \frac{n_A^2 + 2n_A n_B \cosh(2Y_A) + n_B^2}{n_A^2 + 2n_A n_B \cosh(2eY_A) + n_B^2} \notag \\
    &= \frac{(n_A + n_B)^2 \cosh^2(Y_A) - (n_A - n_B)^2 \sinh^2(Y_A)}{(n_A + n_B)^2 \cosh^2(eY_A) - (n_A - n_B)^2 \sinh^2(
    eY_A)}
\end{align}
Since $|e| \leq 1$, $\lambda \geq 1$. If we define $w = \tanh^{-1} \left( \frac{n_A - n_B}{n_A + n_B}\right ) = \frac{1}{2}\ln(\frac{n_A}{ n_B})$, we get the compact expression
\begin{equation}
    \lambda^2 = \frac{\cosh(Y_A - w) \cosh(Y_A + w)}{\cosh(eY_A - w) \cosh(eY_A + w)}
\end{equation}

The mass ratio $\lambda$ also depends on the three quantities $n_A / n_B$, $e$, and $Y_{\rm beam}$. Some example values of $\lambda$ are given in Tables \ref{tab:massratiow0} and \ref{tab:massratiocenterofnucleus} for a variety of different parameters. Note that at $e = 0$ and $w = 0$, we have full stopping of the nucleons, so $\lambda$ is just $\sqrt{s_{\rm NN}} / 2m_{\rm N}$.

\subsection{Rapidity distributions of angular momentum}

The formalism in the previous subsection allows us to determine the rapidity of participant nucleons right after their collision for a given transverse element. Based on the $Y_A'$, $Y_B'$ obtained as functions of $(x, y)$, one can write down a transverse density of the net baryons: 
\begin{equation} \label{eq:dBdxdy}
    \frac{d^2 B}{dxdy} = n_A + n_B
\end{equation} 
However note that the post-collision nucleons from the two sides (the $n_A$ and $n_B$) are at different rapidity $Y_A'$ and $Y_B'$ respectively. One can build a differential rapidity distribution of the net baryon numbers by integrating over the transverse plane in the following way:  
\begin{equation} \label{eq:dBdY}
    \frac{dB}{dY} = \int dx dy \left [ n_A \delta(Y - Y_A') + n_B \delta(Y - Y_B') \right ]
\end{equation} 
where the $Y_A'$ is given in Eq.~\eqref{eq:YA}. Further 
integrating over the full rapidity range gives the net baryon number, which equals $N_{\rm part}$:
\begin{align}
    B &=   \int dx dy \left [ n_A   + n_B   \right]
\end{align}

What we are interested in is the angular momentum distribution just after the collisions, which should be calculated from contributions of all participating nucleons. Using a similar way of counting baryon numbers above, one can write down a transverse density of the angular momentum as: 
\begin{align} \label{eq:dJdxdy}
    \frac{d^2 J_y }{dxdy} = -x \frac{dp_z}{dxdy} = -x \lambda m_{\rm N}   \left [  n_{A} \sinh(Y_{A}') +  n_{B} \sinh(Y_{B}') \right ] 
\end{align}
It shall be emphasized that the $e$-dependence of the above post-collision angular momentum contribution appears in three places: in the post-collision rapidity values $Y_{A}'$ and $Y_{B}'$ as well as in the mass ratio $\lambda$.

By further integrating over transverse plane to take into account contributions from all participant nucleons, one arrives at the following rapidity distribution of initial angular momentum: 
\begin{align} \label{eq:dJdY}
    \frac{dJ_y}{dY} &= -\int dx dy \ x \lambda m_{\rm N} \left [ n_A \sinh(Y_A') \delta(Y - Y_A') + n_B \sinh(Y_B') \delta(Y - Y_B') \right ]
\end{align}
A further integration over the rapidity would give the total angular momentum carried by all participant nucleons: 
\begin{align} \label{eq:Jtot}
   J_y &= -\int dx dy \ x \lambda m_{\rm N} \left [ n_A \sinh(Y_A')   + n_B \sinh(Y_B')   \right ]
\end{align}

\subsection{Implementing fluctuations}
\label{subsec:fluct}

The model can be made more realistic by allowing for fluctuations in the strength of stopping effect. Due to the quantum nature of these scattering processes, the outcome of any given individual collision and thus the resulting rapidity loss would generally take a probabilistic distribution over a range of possible values up to the constraints of exact conservation laws. Such fluctuations can be implemented in our framework by introducing a probability distribution function $f(e)$ for the key elasticity parameter $e$ over the range of $(0, 1)$. (Note negative values of $e$ would correspond to ``bouncing back'' collisions which are not relevant to heavy ion collisions.)    Then the differential baryon number distribution \eqref{eq:dBdY} is modified to
\begin{align} \label{eq:dBdYfe}
    \frac{dB}{dY} &= \sum_{i = A, B} \int_0^1 de f(e) \int dx dy \  n_i \ \delta(Y - Y_i'(e)) \\
    &= \sum_{i = A, B} \int dx dy \  f(e_i(Y)) \  n_i \left| \frac{\partial Y_i'}{\partial e} \right|_{e = e_i(Y)}^{-1}
\end{align}
and \eqref{eq:dJdY} becomes
\begin{align} \label{eq:dJdYfe}
    \frac{dJ_y}{dY} &= -\sum_{i = A, B} \int_0^1 de f(e) \int dx dy \  x  m_{\rm N}  \lambda(x,y,e) \  n_i \sinh(Y_i') \delta(Y - Y_i') \notag \\
    &= -\sum_{i = A, B} \int dx dy \ x m_{\rm N} \lambda(x, y, e_i(Y)) \ f(e_i(Y)) \  n_i \sinh(Y) \left| \frac{\partial Y_i'}{\partial e} \right|_{e = e_i(Y)}^{-1}
\end{align} 
Note in the above, $Y_A'$ and $Y_B'$ as functions of $e$ are given in Eqs.~(\ref{eq:YA},~\ref{eq:YB}) which are both monotonic functions. Therefore the delta functions give at most one root for $e$. The notation $e_{i}(Y)$ (with $i=A,B$) refers to the solutions for $Y_i'(e) = Y$. See details for these solutions as well as for the derivatives in Appendix~\ref{sec:YAB}. 

The next important question is what kind of probability distribution $f(e)$ should be used. To this end, we draw useful ideas from a 2022 paper~\cite{Shen:2022oyg} on initial conditions in which Shen and Schenke present a model for initial baryon stopping based on a probability distribution for fluctuations in the rapidity loss $Y_{\rm loss}$ which they base on the logit-normal distribution~\footnote{The notation is slightly different from Eq. (9) of  {Ref.} \cite{Shen:2022oyg} as we call the independent variable $X$ to avoid confusion with the spatial coordinate $x$.}:
\begin{equation}
    f(X, \sigma) = \frac{1}{\sigma \sqrt{2\pi} X (1 - X)} \exp \left( -\frac{(\mathrm{logit}(X) )^2}{2\sigma^2} \right)
\end{equation}
where $\mathrm{logit}(X) = \ln(X / (1 - X))$. This distribution is supported on $(0, 1)$ and the variable $X$ is related to $Y_{\rm loss}$ by the quadratic equation~\cite{website:shenglauber}
\begin{equation}
    aY_{\rm loss}^2 + bY_{\rm loss} = X
\end{equation}
or
\begin{equation} \label{eq:yloss}
    Y_{\rm loss} = \begin{cases}
        \frac{-b + \sqrt{b^2 + 4aX}}{2a}, & a \neq 0 \\
        Y_{\rm beam} X, & a = 0
    \end{cases}
\end{equation}
The parameters $a,b$ should be chosen to satisfy the constraints that  {$Y_{\rm loss}(X = 0) = 0$, $Y_{\rm loss}(X = 1 / 2) = \langle Y_{\rm loss} \rangle$, and $Y_{\rm loss}(X = 1) = Y_{\rm beam}$}, where $\langle Y_{\rm loss} \rangle$ is the desired mean rapidity loss. As such, one can express $a,b$ as:
\begin{align}
    a &= \frac{2\langle Y_{\rm loss} \rangle - Y_{\rm beam}}{2Y_{\rm beam} \langle Y_{\rm loss} \rangle (Y_{\rm beam} - \langle Y_{\rm loss} \rangle)} \notag \\
    b &= \frac{Y_{\rm beam}^2 - 2\langle Y_{\rm loss} \rangle^2}{2Y_{\rm beam} \langle Y_{\rm loss} \rangle (Y_{\rm beam} - \langle Y_{\rm loss} \rangle)}
\end{align}
unless $Y_{\rm beam} = 2\langle Y_{\rm loss} \rangle$, in which case $a = 0$ and $b = 1 / Y_{\rm beam}$~\cite{website:shenglauber}. The resulting probability distribution for $Y_{\rm loss}$ reads: 
\begin{align}\label{eq:ylosspdf}
    f_{\rm loss}(Y_{\rm loss}, \sigma) &= \frac{2aY_{\rm loss} + b}{\sigma \sqrt{2\pi}  {(aY_{\rm loss}^2 + bY_{\rm loss}) (-aY_{\rm loss}^2 - bY_{\rm loss} + 1)}} \exp \left( -\frac{(\mathrm{logit}(aY_{\rm loss}^2 + bY_{\rm loss}) )^2}{2\sigma^2} \right)
\end{align}
and in terms of $e = 1 - Y_{\rm loss} / Y_{\rm beam}$,
\begin{equation} \label{eq:epdf}
    f(e, \sigma) = Y_{\rm beam} \  f_{\rm loss}((1 - e) Y_{\rm beam}, \sigma) \ \  
\end{equation}  
The above distribution will be used for our calculations, with the width parameter $\sigma$ playing an important role in controlling the magnitude of fluctuations. The desired mean rapidity loss $\langle Y_{\rm loss} \rangle$ can be equivalently quantified by the mean value $\bar{e}$ of the above distribution.

\begin{figure}[!htbp]
    \centering
    \includegraphics[width=0.8\linewidth]{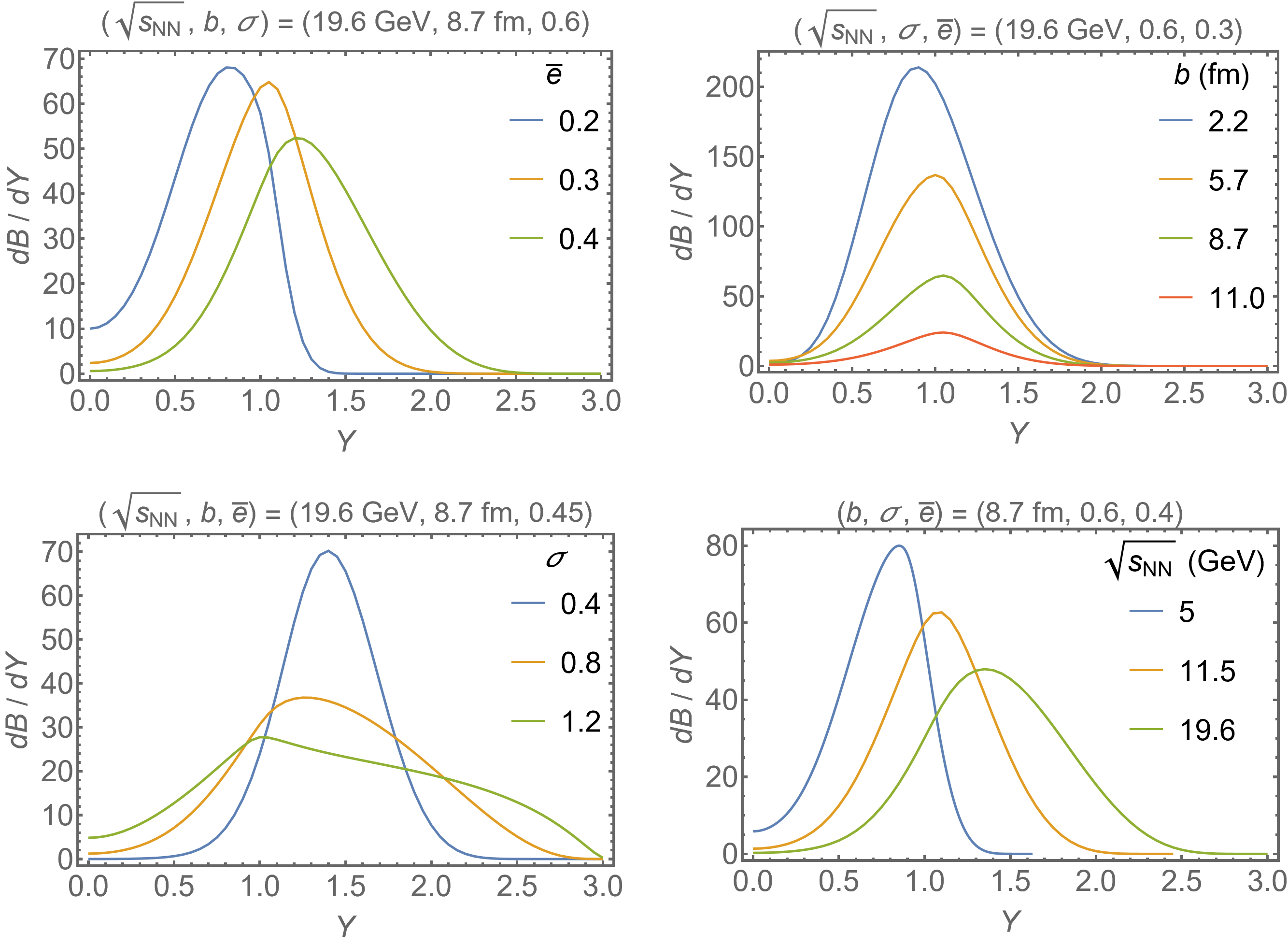}
    \caption{Rapidity distribution of baryon number $dB / dY$ versus rapidity $Y$, for a variety of selected parameter values.}
    \label{fig:dbdy-vs-parameters}
\end{figure}

\begin{figure}[!htbp]
    \centering
    \includegraphics[width=0.8\linewidth]{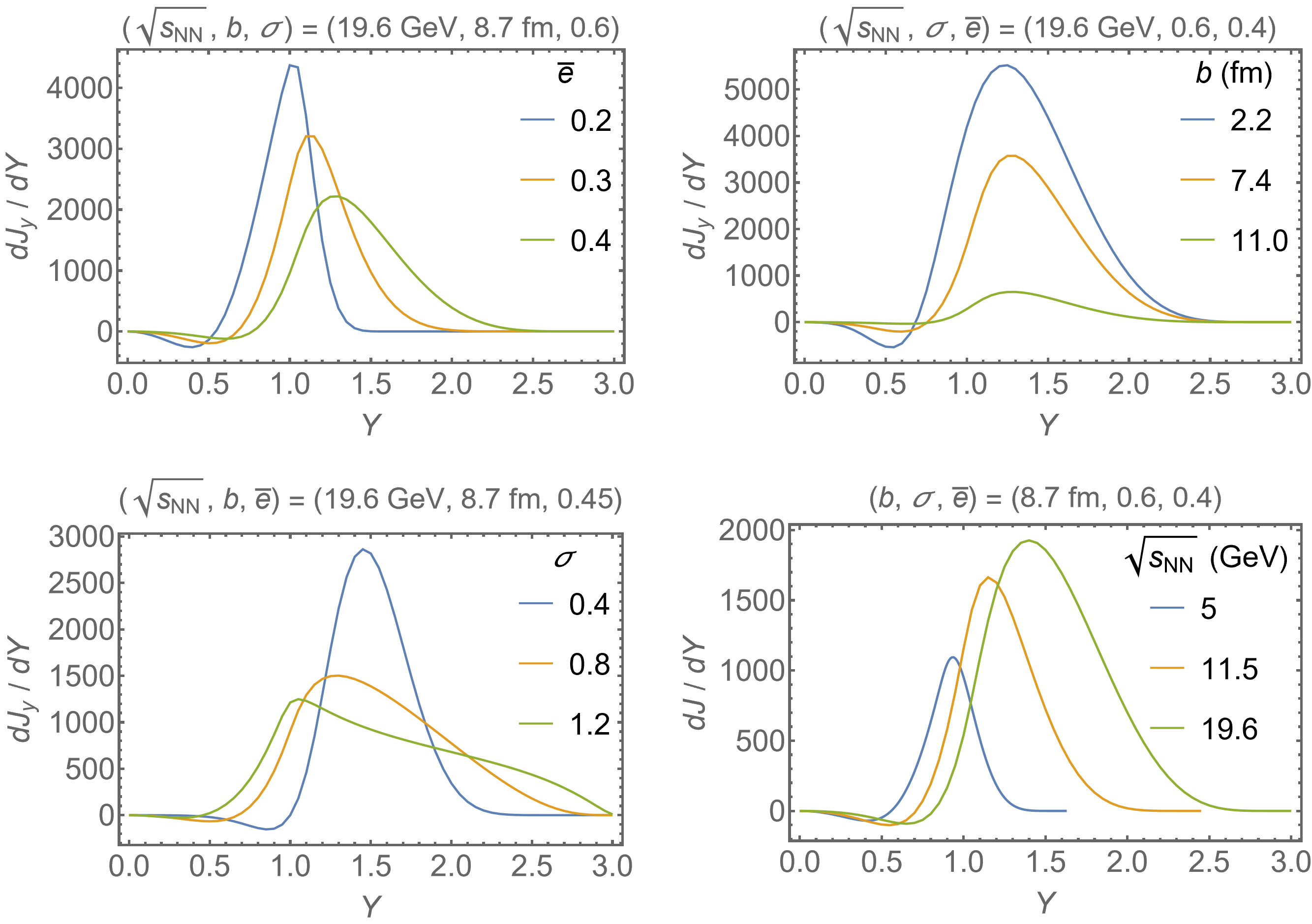}
    \caption{Rapidity distribution of angular momentum $dJ_y / dY$ versus rapidity $Y$, for a variety of selected parameter values.}
    \label{fig:djdy-vs-parameters}
\end{figure}

In short, the present ``Glauber+'' model focuses on describing the rapidity distributions of the participant nucleons (and their contributions to physical quantities like net baryon number and angular momentum) just after the collisions. While the initial nucleons are all at the same $\pm Y_{\rm beam}$, they spread out in rapidity due to (a) different amounts of rapidity loss from collisions at different points on the transverse plane, and (b) fluctuations in the rapidity loss from collisions at a given point on the transverse plane. To give an idea of the resulting distributions, we show some selected results for $dB / dY$ and $dJ_y / dY$ at different parameter values in  {Figs.~}\ref{fig:dbdy-vs-parameters}-\ref{fig:djdy-vs-parameters}. 
Most of the parameter dependence they show is unsurprising \textemdash as $\bar{e}$ increases, the stopping becomes weaker, and the curves move to the larger rapidity region (though they also tend to spread out and flatten). As $b$ increases, the stopping also becomes a little weaker, as there is generally more of a mismatch between the densities of the target and projectile, leading to a larger first term in \eqref{eq:YA}. The main effect of a larger $b$, though, is to decrease the overall height of the curve as less matter is in the collision zone. Increasing $\sigma$ broadens the curve, but also makes it less symmetric as the non-Gaussian behavior of $f(e)$ becomes more important. Finally, increasing $\sqrt{s_{\rm NN}}$ means there is less stopping in an absolute sense as the curve shifts to the right, and the overall scale of both 
$dB / dY$ and  {$dJ_y / dY$} grow with $\sqrt{s_{\rm NN}}$ as they should. 

\section{Results}  \label{sec:results}

\subsection{Parameter calibration} \label{subsec:calib}

With the model introduced in the previous section, we would now use relevant experimental information to help calibrate the key model parameters. These include the average value of the elasticity parameter $\bar{e}$ and the width $\sigma$ that controls its fluctuations. With the assumption that the baryon stopping at mid-rapidity is predominantly due to the initial collisions, we could compare the model results computed a wide variety of $\bar{e}$ and $\sigma$ for $dB / dY$ with experimental measurements for net baryon number at mid-rapidity. One complication is that only the net-proton number is experimentally measured and thus an estimate of the net-baryon-to-net-proton ratio, or baryon correction factor, is required. The details of estimating this factor is discussed in Appendix\autoref{sec:baryonfactor}.

Let us now discuss some details of this calculation. As described above, the nuclei are described by a Woods-Saxon density, which is integrated over the $z$-axis to give a thickness function, and the thickness function is stored as a table of values $(r_\perp, T(r_\perp))$ with step size $\delta r_\perp = 10^{-5} \textrm{ MeV}^{-1}$ or about $0.002$ fm. 
Integration is done numerically in Python with the trapezoidal method, using evenly spaced intervals in $x, y$ of size $\delta x = \delta y = 0.0005 \textrm{ MeV}^{-1} \approx 0.1 \textrm{ fm}$. The results are improved with first-order Romberg quadrature. Calculations are done for a broad range of beam energy $\sqrt{s_{\rm NN}}$ and  {centrality}, including:
$\sqrt{s_{\rm NN}} \in \{$2, 2.42, 3, 5, 7.7, 11.5, 14.5, 19.6, 27, 39, 54.4, 62.4, 200$\}$ GeV and centrality $\in \{$0-5, 5-10, 10-20, 20-30, 30-40, 40-50, 50-60, 60-70, 70-80$\} \%$. 

 {Since the Glauber calculations depend on the impact parameter $b$, a relation between $b$ and centrality is necessary when comparing to experimental data. This relation has been calculated using Monte Carlo Glauber simulations in the literature. For example, Table XX of  {Ref.} \cite{Loizides:2017ack} lists minimum and maximum impact parameter values $b_{min}$ and $b_{max}$ for different centrality bins in Au-Au collisions at $\sqrt{s_{\rm NN}} = 200$ GeV. One can calculate an average $\hat{b}$ for each bin as}

\begin{equation}
    \hat{b} = \frac{\int_{b_{\rm min}}^{b_{\rm max}} b^2 db}{\int_{b_{\rm min}}^{b_{\rm max}} b db} = \frac{2}{3} \frac{b_{\rm max}^3 - b_{\rm min}^3}{b_{\rm max}^2 - b_{\rm min}^2}
\end{equation} 
 {Thus, for the 0-5\% bin, $\hat{b} = 2.21$ fm. All of our calculations for a given centrality bin were done using the corresponding $\hat{b}$. We note this approach requires that the correspondence between ``geometric'' quantities such as impact parameter with centrality bins is not too sensitive to the beam energy $\sqrt{s_{\rm NN}}$. Monte Carlo Glauber calculations seem to suggest that this is the case, at least down to BES-I energies: see e.g. Tables 2-5 of  {Ref.} \cite{Ray:2007av}, Tables II-III of  {Ref.} \cite{STAR:2012och}, Table II of  {Ref.} \cite{STAR:2008ftz}. }

\begin{figure}[!htbp]
    \centering
    \subfloat[]{
        \includegraphics[width=0.4\textwidth]{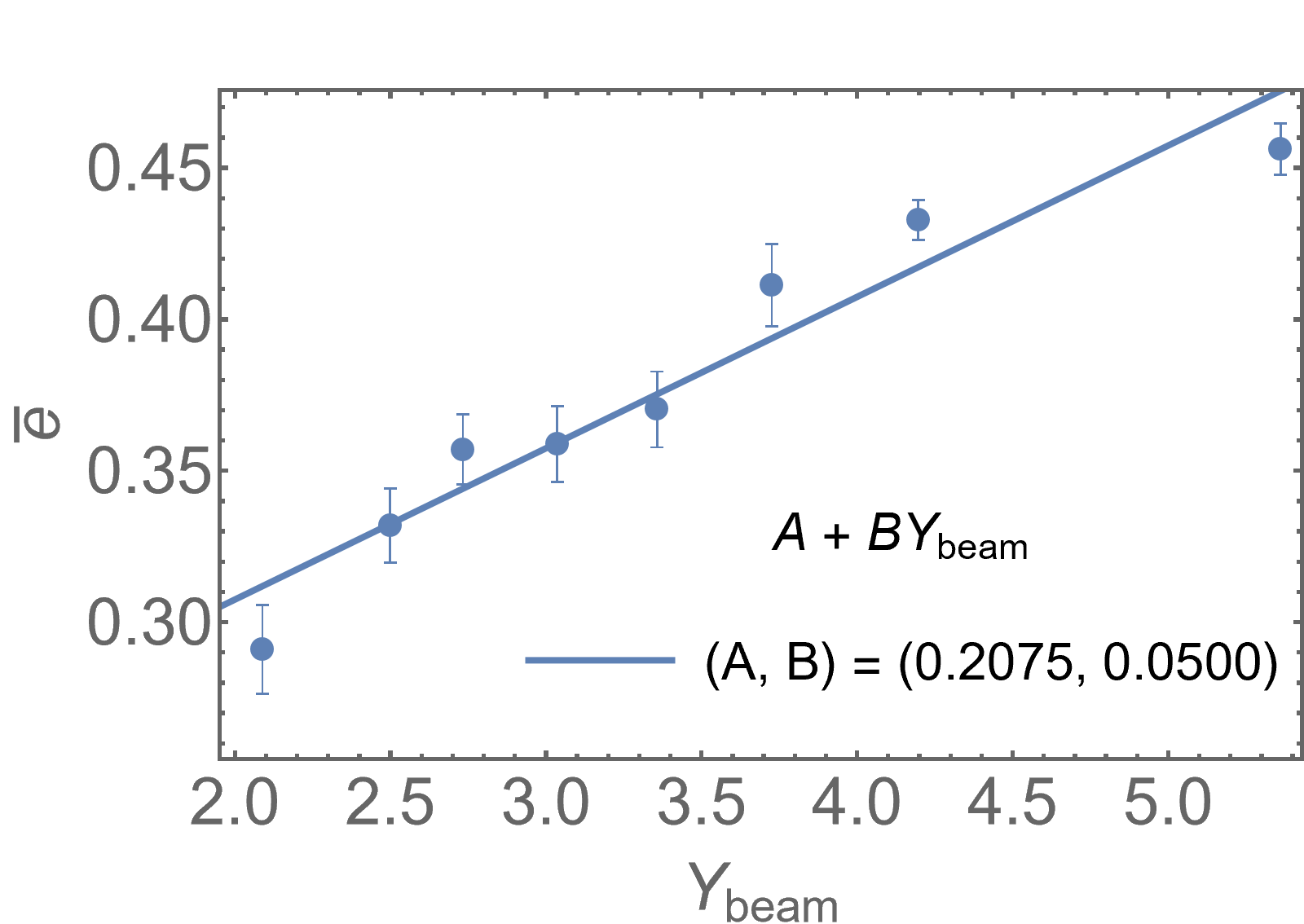}
        \label{fig:efitplot}
    }
    \subfloat[]{
        \includegraphics[width=0.4\textwidth]{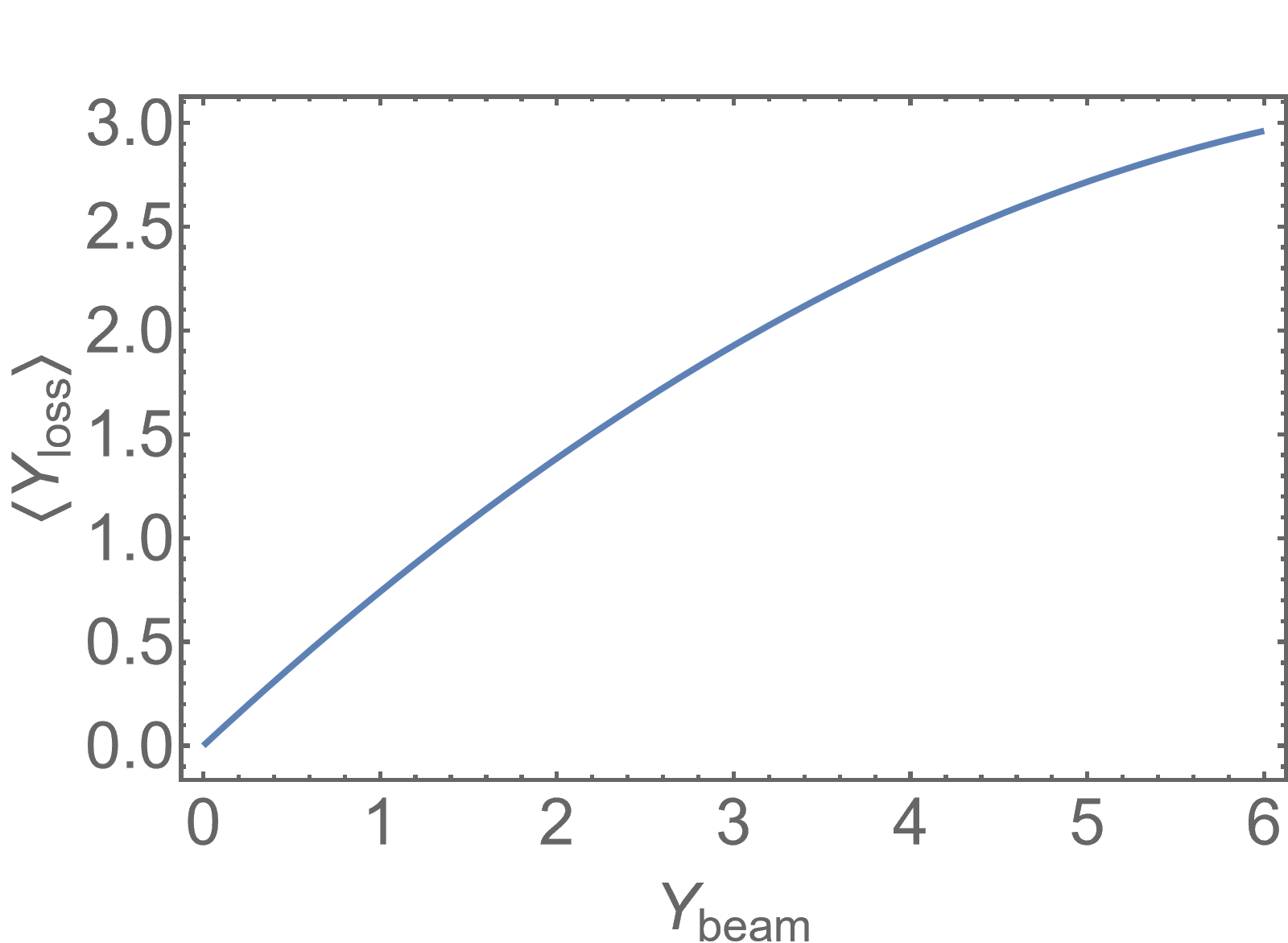}
        \label{fig:ylossplot}
    }
    \caption{
         {(a)} The extracted optimal $\bar{e}$ values at several beam energies where STAR measurements are available. A linear fit of this dependence is shown as the straight line. See text for details.  {(b) The corresponding average rapidity loss $\langle Y_{\rm loss} \rangle = (1 - \bar{e}) Y_{\rm beam}$, computed with $\bar{e}$ from the fit in (a). }
    }
    \label{fig:efitandylossplot}
\end{figure}

For each given beam energy and centrality, calculations are done for different choices of the two model parameters, covering the range of $\bar{e} \in [0.2, 0.55]$ (though the upper bound was not reached for most $\sqrt{s_{\rm NN}}$) and $\sigma \in [0.4, 1.2]$.
The most optimal values are chosen for $(\bar{e}, \sigma)$ at each energy by comparison with measurements. Experimental net-proton data are available for 7.7, 11.5, 19.6, 27, and 39 GeV from  {Ref.} \cite{STAR:2017sal}, for 14.5 GeV from  {Ref.} \cite{STAR:2019vcp}, and for 62.4 and 200 GeV from  {Ref.} \cite{STAR:2008med}. With the baryon correction factor discussed in Appendix\autoref{sec:baryonfactor}, we can reasonably estimate the net-baryon number in mid-rapidity as a function of beam energy and centrality. For a given parameter set $(\sqrt{s_{\rm NN}}, b, \sigma)$, then, we can adjust $\bar{e}$ until the predicted $dB / dY$ best matches the estimated experimental $dB / dY$ (at mid-rapidity per kinematic cut from experimental analysis). For simplicity we assume a uniform $\sigma$ value for all energies, for which the best choice is found to be $\sigma = 1.0$. At each $\sqrt{s_{\rm NN}}$, a mean of the resulting $\bar{e}$ values is taken over all centrality bins.  
The resulting $\bar{e}$ versus beam rapidity $Y_{\rm beam}$ is shown in Fig.~\ref{fig:efitplot}. This dependence can be used to extrapolate $\bar{e}$ to other energies of interest by performing a linear fitting analysis for $\bar{e}(Y_{\rm beam})$. 
The result, $\bar{e}(Y_{\rm beam}) =  {0.2075 + 0.0500 Y_{\rm beam}}$ is shown as the straight line in  {Fig.~\ref{fig:efitplot}. The corresponding average rapidity loss $\langle Y_{\rm loss} \rangle = (1 - \bar{e}) Y_{\rm beam}$ on $Y_{\rm beam}$, using the fitted $\bar{e}(Y_{\rm beam})$, is shown in Fig.~\ref{fig:ylossplot}.}

\begin{figure}[!htbp]
    \centering
    \includegraphics[width=0.8\textwidth]{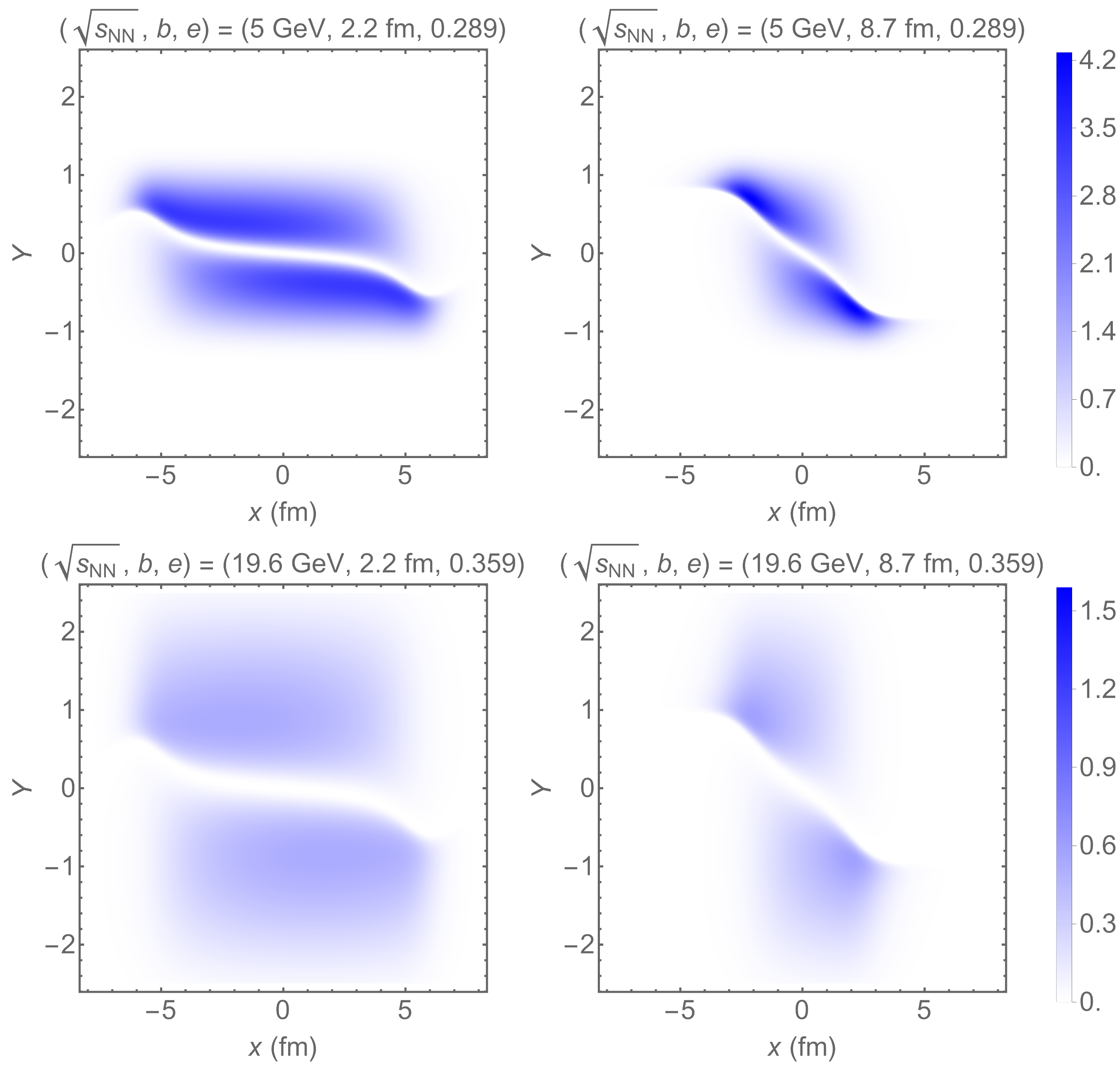}
    \caption{ {The triply-differential baryon number density $d^3B / dxdydY$ (in $\textrm{fm}^{-2}$) evaluated at $y = 0$ for a few choices of $\sqrt{s_{\rm NN}}$ and $b$. Calibrated $\bar{e}$ values are used, and $\sigma = 1.0$. Note the reflection symmetry of the density with respect to $(x, Y)$.}}
    \label{fig:dbdxdydY}
\end{figure}

 {It would be interesting to have an intuitive picture of the baryon stopping immediately after the collision, as determined by the triply-differential quantity $d^3B / dxdydY$. We show its 2D density plot on the $(x, Y) $ plane with $y = 0$ in Fig.~\ref{fig:dbdxdydY} at two beam energies and two impact parameters, using calibrated $\bar{e}$ and $\sigma$. Some qualitative features are clear from the plots, such as: the symmetry of the density under the reflection $(x, Y) \rightarrow (-x, -Y)$, the decrease in stopping with increasing beam energy, the decrease in stopping toward the edges of collision zone, etc.}

\subsection{Results for initial angular momentum} 
\label{subsec:results}
\begin{figure}[!htbp]
    \centering
    \includegraphics[width=0.8\linewidth]{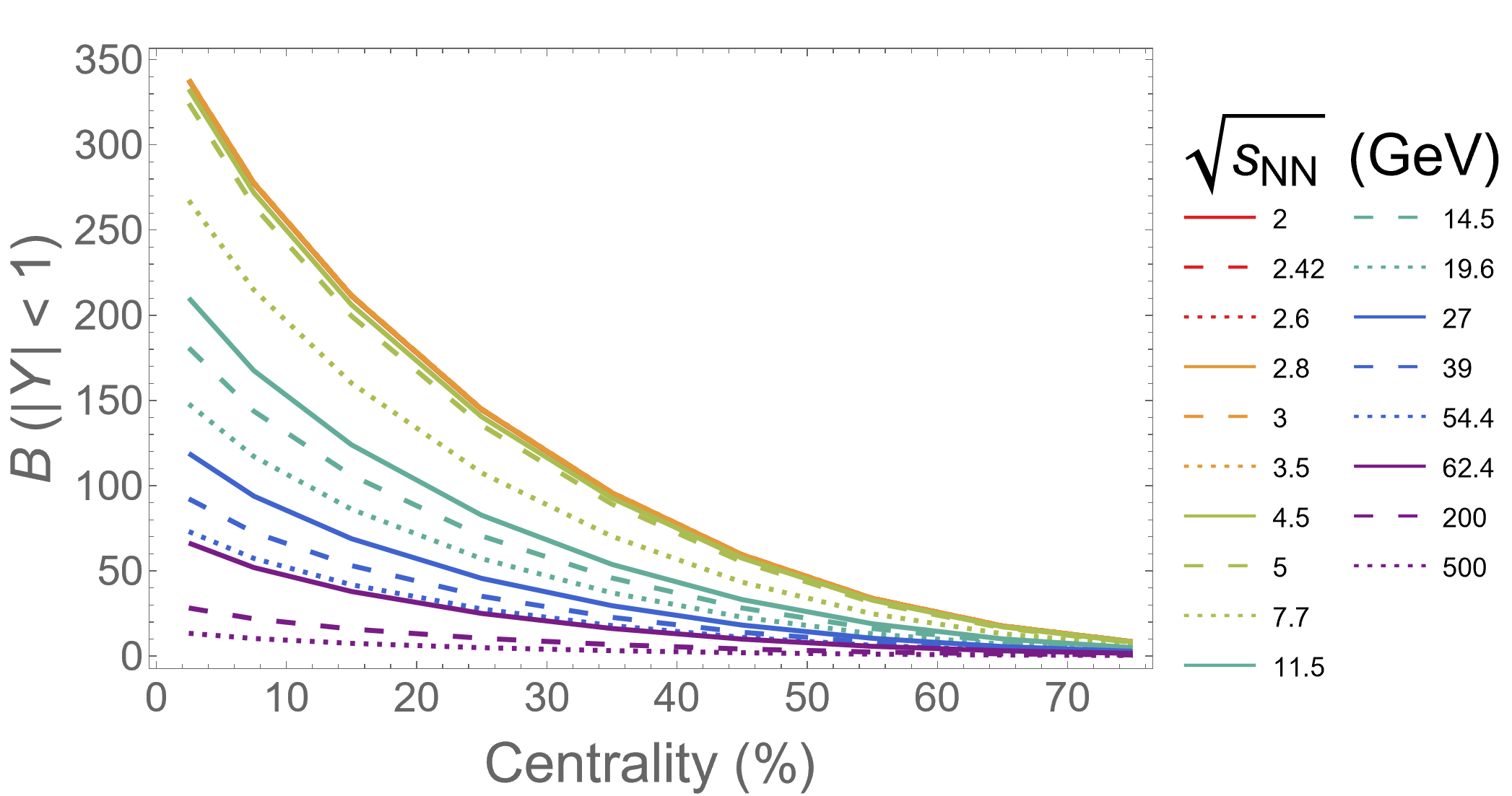}
    \caption{The net baryon number $B$ in mid-rapidity $|Y| < 1$ versus centrality at different beam energies $\sqrt{s_{\rm NN}}$.  {Note that some of the lowest energies produce visually indistinguishable curves here.}}
    \label{fig:dbdy-final}
\end{figure}

\begin{figure}[!htbp]
    \centering
    \includegraphics[width=0.8\linewidth]{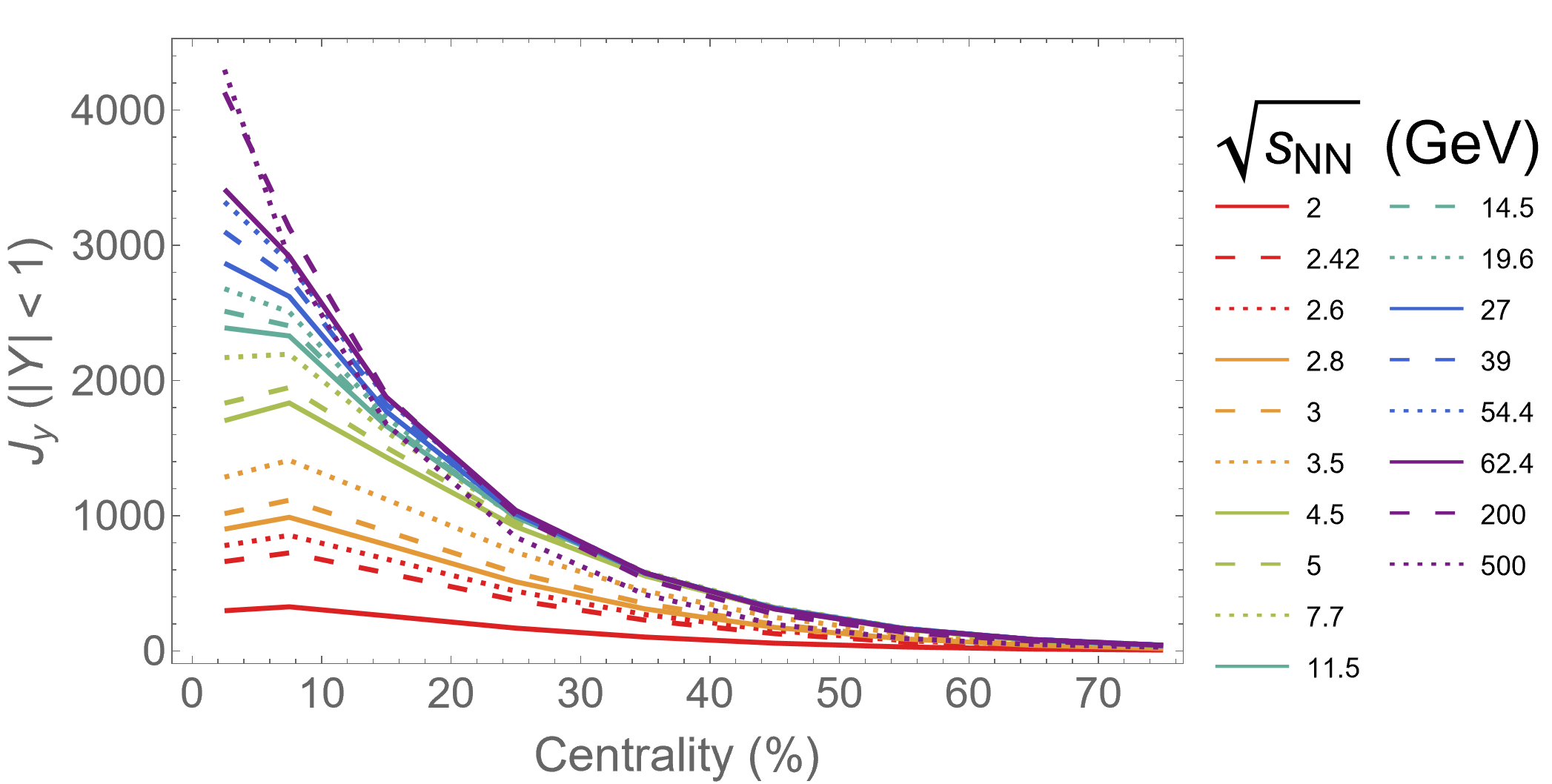}
    \caption{The angular momentum $J_y$ in mid-rapidity $|Y| < 1$ versus centrality at different beam energies $\sqrt{s_{\rm NN}}$.  {Note that some of the curves become visually indistinguishable, especially for some of the highest energies and in the peripheral region. }}
    \label{fig:djdy-final}
\end{figure}

Finally, we are ready to compute the quantities of most interest: the initial net baryon number and angular momentum in the mid-rapidity fireball. This can be obtained by integrating Eqs.~(\ref{eq:dBdY},~\ref{eq:dJdY}) over the rapidity interval $Y\in (-1,1)$. These results are shown in Figs.~\ref{fig:dbdy-final} and \ref{fig:djdy-final}. 

For net baryon number, the interpretation is straightforward as it declines in a monotonic fashion with respect to beam energy $\sqrt{s_{\rm NN}}$ as well as to centrality, as expected. However, $J_y$ is more complex. Changing from central to peripheral collisions, the orbital angular momentum first grows with increasing $b$ but then decreases due to the decline in the number of participant nucleons in the collision zone. The competition between these effects gives rise to a peak for 5-10\% centrality for $\sqrt{s_{\rm NN}} \leq 7.7$ GeV. However, at higher $\sqrt{s_{\rm NN}}$ there is a monotonic decline with respect to $b$ for the centrality bins we consider here. As for the $\sqrt{s_{\rm NN}}$-dependence, $J_y$ at mid-rapidity first grows with beam energy, for the obvious reason of increasing $z$-momentum of nucleons. However, only for collisions with 0-5\% centrality is the dependence strictly monotonic.  Generally, for other centralities, $J_y$ at mid-rapidity peaks at an intermediate $\sqrt{s_{\rm NN}}$ then decreases. This peak can be as low as $7.7$ GeV, though for higher energies the value changes little as can be seen from Fig.~\ref{fig:djdy-final}.  This non-monotonicity can be attributed to the weakening of the stopping effect as $\sqrt{s_{\rm NN}}$ grows.

\begin{figure}[!htbp]
    \centering
    \includegraphics[width=0.8\linewidth]{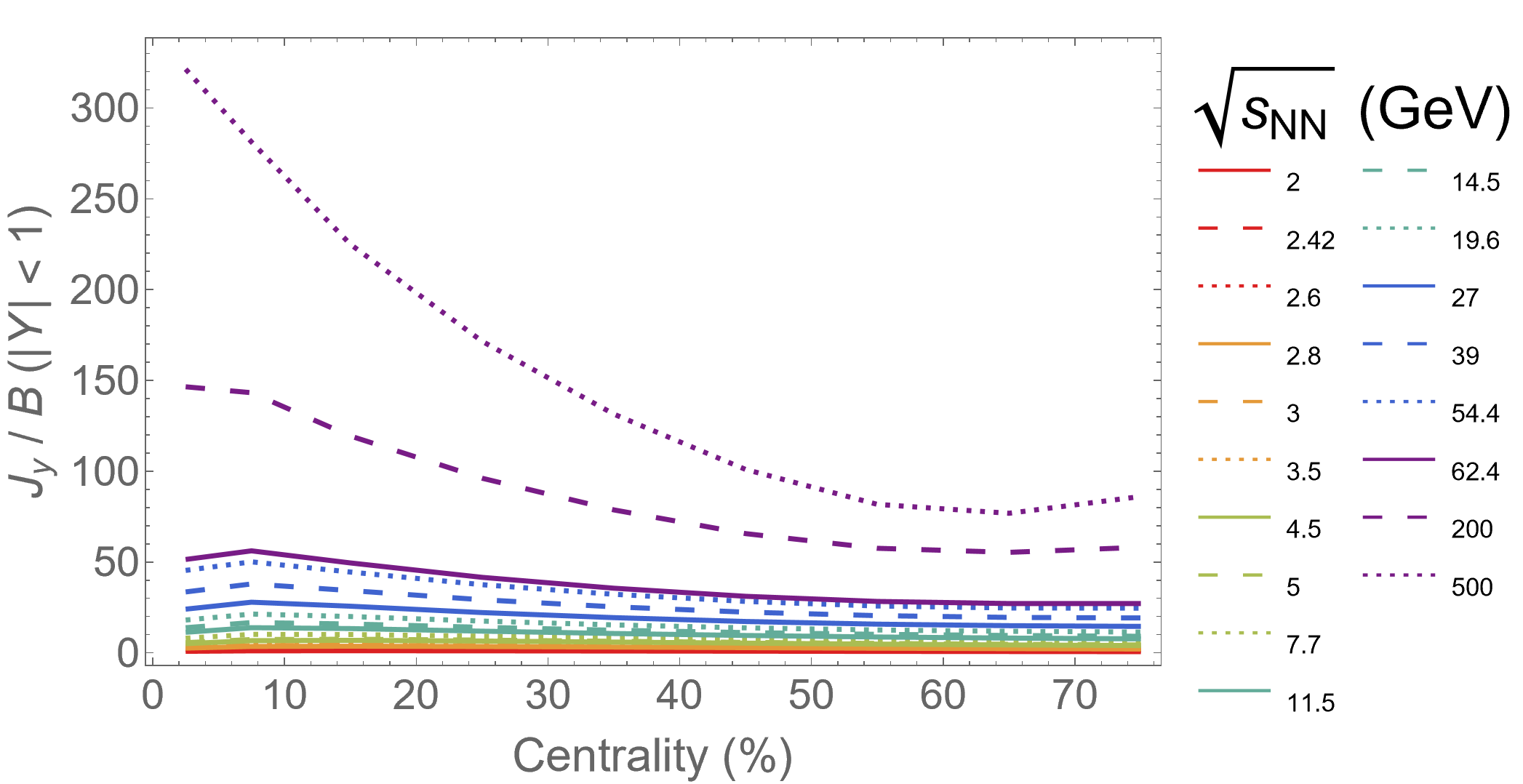}
    \caption{The ratio $J_y / B$ in mid-rapidity $|Y| < 1$ versus centrality at different beam energy $\sqrt{s_{\rm NN}}$.
     {Note that some of the lowest energies produce nearly indistinguishable curves here.} }
    \label{fig:j-over-b-final}
\end{figure}

One interesting quantity to look at is the ratio between the initial angular momentum and net baryon number, $J_y / B$, in which the net baryon number serves as a calibration of the stopping effect. Fig.~\ref{fig:j-over-b-final} shows that the angular momentum per net baryon number at mid-rapidity grows monotonically with beam energy, and that the peak at 5-10\% centrality persists up to at least $62.4$ GeV.

\begin{figure}[!htbp]
    \centering
    \includegraphics[width=0.7\linewidth]{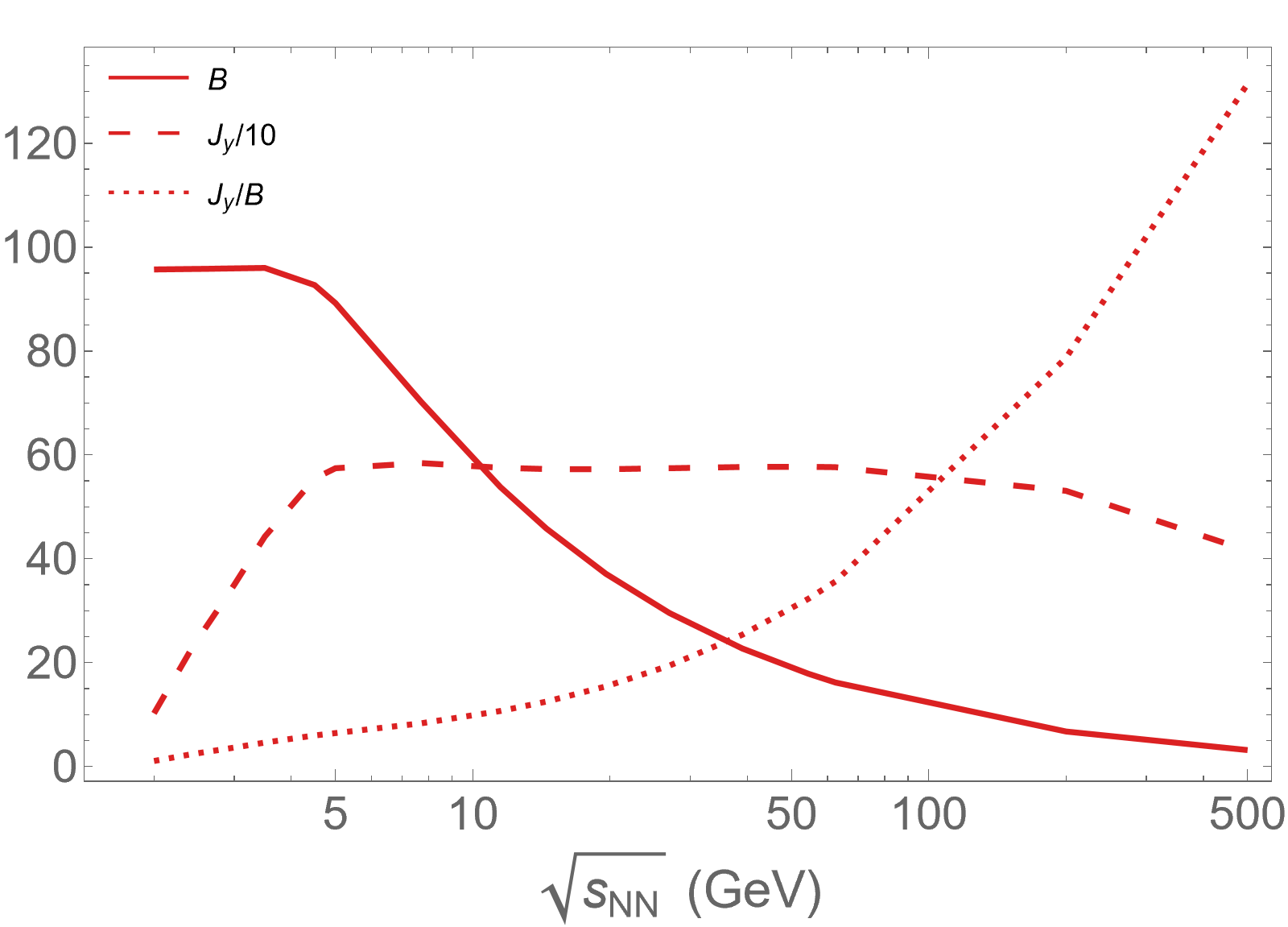}
    \caption{The quantities $B$, $J_y$, and $J_y / B$ in mid-rapidity ($|Y| < 1$) versus beam energy $\sqrt{s_{\rm NN}}$ for 30-40\% centrality collisions.}
    \label{fig:all-vs-s-final}
\end{figure}

Lastly, we investigate the beam energy dependence of relevant quantities for a fixed centrality range of interest. Here we choose to focus on the 30-40\% centrality which is most relevant to the global polarization measurements in, e.g.,  {Refs.} \cite{STAR:2021beb,HADES:2022enx}.  {Fig.~}\ref{fig:all-vs-s-final} shows quantities $B$, $J_y$, and $J_y / B$ in mid-rapidity ($|Y| < 1$) versus beam energy. The net baryon number $B$ decreases with increasing beam energy as expected. The angular momentum $J_y$, on the other hand, shows a more complex pattern. It first increases and seems to plateau from about 5 to 50 GeV before declining again. This nontrivial dependence with increasing $\sqrt{s_{\rm NN}}$ is due to the two competing effects, namely the growth of total angular momentum versus the weakening of stopping at mid-rapidity. Not surprisingly, the ratio $J_y / B$ essentially normalizes away the stopping effect and demonstrates a monotonic increase with beam energy.

\begin{figure}[!htbp]
    \centering
    \includegraphics[width=0.7\linewidth]{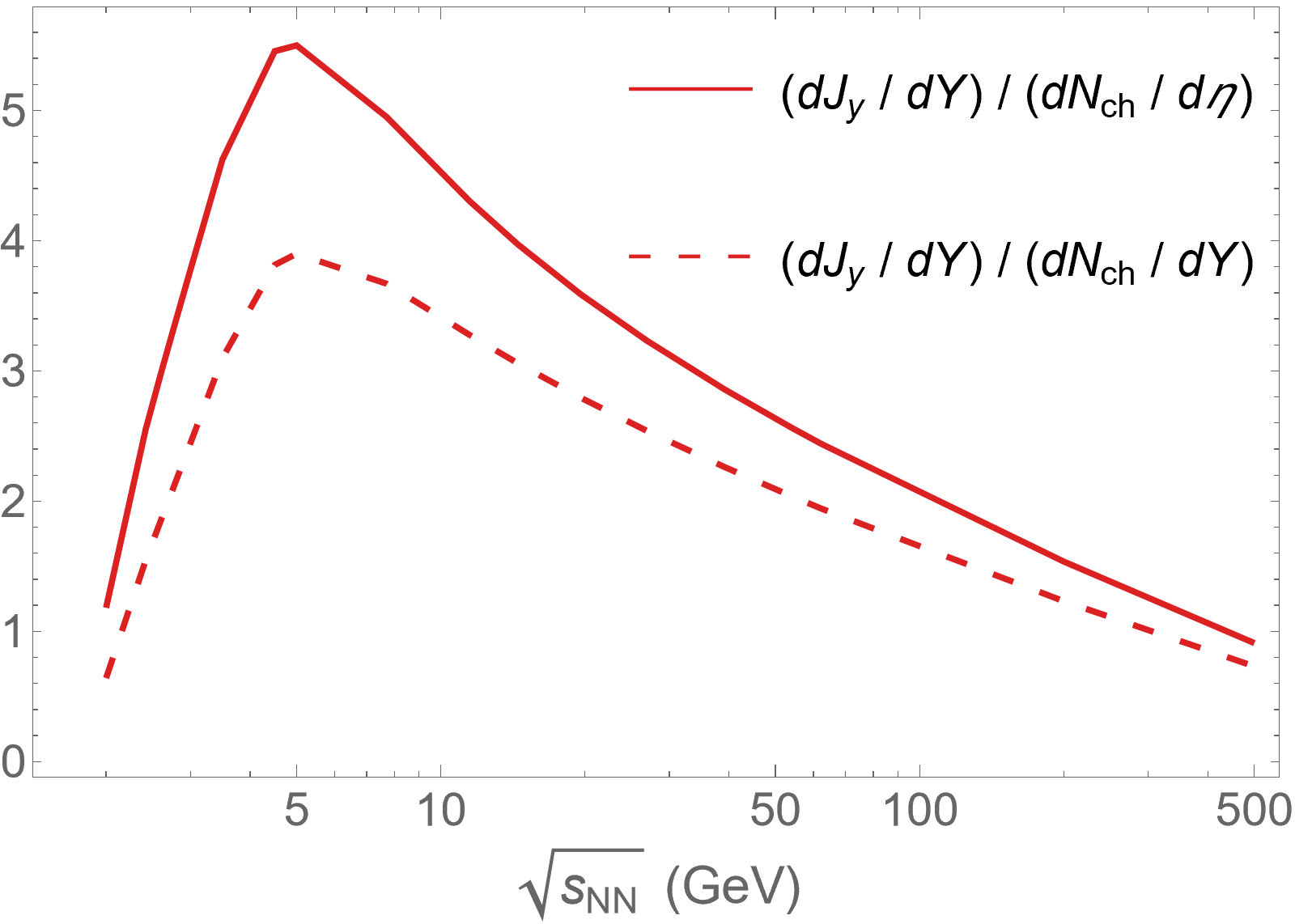}
    \caption{The dependence of ratios $\frac{dJ / dY}{dN_{ch} / d\eta}$ (solid) and $\frac{dJ / dY}{dN_{ch} / dY}$ (dashed) in mid-rapidity on beam energy $\sqrt{s_{\rm NN}}$ for 30-40\% centrality collisions. See text for details.} 
    \label{fig:j-over-nch-vs-s-final}
\end{figure}

Given that we are interested in possible global spin polarization driven by angular momentum, a valuable quantity to examine would be the {\em angular momentum per final state hadron}. Note that the fireball angular momentum is conserved from initial state to final state~\cite{Jiang:2016woz} and that the measured charged hadron multiplicity serves as a good proxy for the total number of final state hadrons. Therefore one can construct a useful quantitative measure as $(dJ_y / dY) / (dN_{ch} / d\eta)$ in mid-rapidity region, where $dJ_y / dY$ is the initial angular momentum computed from our model and $dN_{ch} / d\eta$ is the final state charged multiplicity from experimental measurements. For the latter, we use an empirical formula derived from fitting available data, given in  {Ref.} \cite{NathMishra:2015qrc} as
\begin{align}
    \frac{dN_{ch}}{d\eta} &= \frac{N_{\rm part}}{2} \ f(s) \  g(N_{\rm part}) \\
    f(s) &= 0.0147 (\ln(s))^2 + 0.6 \\
    g(N_{\rm part}) &= 1 + 0.095 N_{\rm part}^{1 / 3}
\end{align}
where $N_{\rm part}$ can then be calculated in our model.  {Fig.~}\ref{fig:j-over-nch-vs-s-final} (solid curve) shows the results. One sees that with increasing beam energy, the ratio first rises quickly from the threshold and peaks somewhat sharply around 5 GeV, before gradually dropping off at higher $\sqrt{s_{\rm NN}}$. 

One minor issue in the above ratio is that the two differential quantities are with respect to different variables, $Y$ and $\eta$ respectively. One may be curious about the impact of a Jacobian factor $d \eta / dY$ that accounts for the difference. This factor has been estimated at mid-rapidity by the PHENIX Collaboration: see Table V of  {Ref.} \cite{PHENIX:2015tbb} where estimates are given for $d\eta / dY$ at several energies. These values are well-described by a fitting function $A + Be^{-Y_{\rm beam}}$, where $(A, B) \approx (1.243, 0.857)$. 
Assuming the extrapolation of this function to the beam energy region of our interest, we can now make an estimate of $\frac{dJ_y / dY}{dN_{ch} / dY}$, with the results shown in Fig.~\ref{fig:j-over-nch-vs-s-final} (dashed curve). We can see that both the qualitative trend and the quantitative features remain largely unchanged, with a visible peak still around $\sqrt{s_{\rm NN}} \approx 5$ GeV.

\section{Conclusion} \label{sec:conclusion}

In summary, we've developed an optical Glauber+ model that calculates the initial baryon stopping and angular momentum in heavy ion collisions over a broad range of collisional beam energies, from $\mathcal{O}(1)$ GeV to $\mathcal{O}(1000)$ GeV. By extending the traditional Glauber model to the early moment just after initial impacts and implementing generic (yet mandatory) requirements of exact conservation laws, this model predicts the rapidity distributions of initial net baryon number and angular momentum after calibrating model parameters with available experimental data. A natural next step will be performing Monte Carlo simulations based on the current Glauber+ model and applying the obtained initial conditions for dynamical simulations of relevant observables such as spin polarization and alignment.

A key issue at stake, which is a motivating factor for the present study, is the beam energy dependence of the global spin polarization signal, especially its trend toward the very low energy region. If one assumes that the global spin polarization is dominantly driven by the fireball's angular momentum, then our findings strongly suggest that the peak of this phenomenon (and any other phenomena induced by angular momentum) would be around $\sqrt{s_{\rm NN}}\approx 5$ GeV and is highly unlikely to be very close to the threshold energy at $\sqrt{s_{\rm NN}} = 2 m_{\rm N} $. The current experimental data, indicating a rising trend toward energy as low as $2.4$ GeV, would then have to suggest alternative interpretations of the observed global spin polarization other than the system's angular momentum at extremely low energy, if ultimately validated to be the case. A number of potential sources for such alternative contributions~\cite{Xie:2019wxz,Guo:2019joy,Guo:2019mgh,Sheng:2022wsy,Ivanov:2019ern,Ivanov:2022ble,Sung:2024vyc,Sun:2021nsg} could include, e.g., hadronic mean field effects, angular momentum transport from spectators, strong magnetic fields, intrinsic polarization effects in hadronic processes that could dominate strangeness production in low energy collisions, etc.

\section*{Acknowledgments}

The authors thank N. Xu and H. Zhang for useful discussions. The authors acknowledge support by the U.S. NSF under Grant No.~PHY-2209183, ~PHY-2514992, and by the U.S. DOE through ExoHad Topical Collaboration under Contract No.~DE-SC0023598. A.A. was also supported by the U.S. DOE under Grant No. DE-FG02-87ER40365. H.M. was supported by the U.S. NSF via the REU program at Indiana University (Grant No.~PHY-2150234). Computation for this research was performed at the Big Red 200 supercomputer system which was supported in part by Lilly Endowment, Inc., through its support for the Indiana University Pervasive Technology Institute.

\appendix

\section{Additional details for rapidity loss calculations} 
\label{sec:YAB}

\subsection{Local COM frame} 

One useful point to note is that the two terms in Eq.~\eqref{eq:YA} could be understood in a more transparent way by boosting to the {\em local COM frame} defined by the two colliding bunches of nucleons within the area element under consideration. Let $\zeta$ be the rapidity for a boost from the global COM frame to this local COM frame, and let overbars represent quantities in the latter. Then
\begin{align}
    0 &= \frac{d\bar{p}_z}{m_{\rm N} dxdy} = n_A \sinh(Y_A + \zeta) + n_B \sinh(-Y_A + \zeta) \\
    &= (n_A - n_B) \sinh(Y_A) \cosh(\zeta) + (n_A + n_B) \cosh(Y_A) \sinh(\zeta) \\
    \implies \zeta &= -\tanh^{-1} \left( \frac{n_A - n_B}{n_A + n_B} \tanh(Y_A) \right)
\end{align}
Since boosts in the $z$-direction are just shifts in rapidity, they leave $e$ unchanged, so we can still take $\bar{Y}_B' = \bar{Y}_A' - 2eY_A$ if $Y_A$ is taken to be the beam rapidity in the global COM frame. Then conservation of $z$-momentum becomes
\begin{align}
    0 &= n_A \sinh(\bar{Y}_A') + n_B \sinh(\bar{Y}_B') \notag \\
    &= (n_A + n_B \cosh(2eY_A)) \sinh(\bar{Y}_A') - n_B \sinh(2eY_A) \cosh(\bar{Y}_A')
\end{align}
and $\bar{Y}_A' = u$. We see that the two terms in \eqref{eq:YA} are just the rapidity needed to boost to the local COM frame and the final-state rapidity of an $A$ nucleon in that frame. Note that if $n_A = n_B$ at a point in the transverse plane, then the global and local COM frames are the same there, and
\begin{equation}
    u = \tanh^{-1} \left( \frac{\sinh(2eY_A)}{1 + \cosh(2eY_A)} \right) = eY_A = e\bar{Y}_A
\end{equation}

\subsection{Relations between $e$ and $Y_{A,B}'$} 

Another useful point to note is that for given $n_A$ and $n_B$, we can also invert Eq.~\eqref{eq:YA} to determine the $e$ value that corresponds to a given $Y'$ in the global COM frame. First, calculate
\begin{equation}
    \bar{Y}_A' = Y_A' - \tanh^{-1} \left( \frac{n_A - n_B}{n_A + n_B} \tanh(Y_A) \right)
\end{equation}
and then from
\begin{equation}
    \tanh(\bar{Y}_A') = \frac{n_B \sinh(2eY_A)}{n_A + n_B \cosh(2eY_A)}
\end{equation}
a short calculation gives
\begin{equation} \label{eq:efromY1}
    \sinh(2eY_A - \bar{Y}_A') = \frac{n_A}{n_B} \sinh(\bar{Y}_A')
\end{equation}
and
\begin{equation} \label{efromY}
    e_A(Y_A) \equiv e = \frac{\bar{Y}_A' + \sinh^{-1} \left( \frac{n_A}{n_B} \sinh(\bar{Y_A}') \right)}{2Y_A}
\end{equation}
Similarly, since $\bar{Y}_B' = \bar{Y}_A' - 2eY_A$, $-\sinh(\bar{Y}_B') = (n_A / n_B) \sinh(\bar{Y}_B' + 2eY_A)$ and
\begin{equation} \label{eq:efromYB}
    e_B(Y_B) = \frac{\bar{Y}_B' + \sinh^{-1} \left( \frac{n_B}{n_A} \sinh(\bar{Y_B}') \right)}{-2Y_A}
\end{equation}
which defines the functions $e_{A, B}(Y)$. One last result that will be useful is the derivative of $Y_A'$ (or $Y_B'$) with respect to $e$:
\begin{align} \label{eq:dYde}
    \frac{\partial Y_A'}{\partial e} &= 2n_B Y_A \frac{n_B + n_A \cosh(2eY_A)}{n_A^2 + 2n_A n_B \cosh(2eY_A) + n_B^2} \notag \\
    &= 2Y_A \frac{1 + \frac{n_A}{n_B} \cosh(2eY_A)}{1 + 2 \frac{n_A}{n_B} \cosh(2eY_A) + \frac{n_A^2}{n_B^2}} \\
    \frac{\partial Y_B'}{\partial e} &= -2Y_A \frac{1 + \frac{n_B}{n_A} \cosh(2eY_A)}{1 + 2 \frac{n_B}{n_A} \cosh(2eY_A) + \frac{n_B^2}{n_A^2}}
\end{align}
Note that neither of these change sign (or become zero), so $Y_{A, B}'$ are monotone functions of $e$.

\section{Estimating the baryon number correction factor} \label{sec:baryonfactor}

Here we discuss a method for converting net-proton number, which is experimentally measured, to net-baryon number that we calculate in the present model. The key issue is to get a reasonable estimate of contributions from (anti)neutrons to the net-baryon number, so the problem then reduces to estimating their yields. The most straightforward way to do this is based on thermal models. 
In  {Ref.} \cite{STAR:2019sjh}, the relativistic thermal formula for the yield of a hadron species $i$ with baryon number $B_i$, strangeness $S_i$ and electric charge $Q_i$ is given as
\begin{equation}
    N_i = \frac{g_i V}{\pi^2} m_i^2 T K_2(m / T) \exp(\mu_i / T)
\end{equation}
where the species' chemical potential is $\mu_i = B_i \mu_B + S_i \mu_S + Q_i \mu_Q$, $g_i$ is the degeneracy of that species, $T$ is the chemical freeze-out temperature, and $V$ is the freeze-out volume from which the hadron species is thought to be produced. The relevant quantum numbers are given in Table \ref{tab:qn}.
\begin{table}[!hbt]
    \begin{tabular*}{0.5\textwidth}{@{\extracolsep{\fill}}|ccccccccc|}
    \hline
    Species & $p$ & $\bar{p}$ & $n$ & $\bar{n}$ & $d$ & $\bar{d}$ & $\pi^+$ & $\pi^-$ \\
    \hline
    $B$ & 1 & -1 & 1 & -1 & 2 & -2 & 0 & 0 \\
    \hline
    $Q$ & 1 & -1 & 0 & 0 & 1 & -1 & 1 & -1 \\
    \hline
    \end{tabular*}
    \caption{Baryon numbers and electric charges for relevant hadrons (here $d$ means the deuteron).}
    \label{tab:qn}
\end{table}

Let $p, n$ ($\bar{p}, \bar{n}$) stand for the (anti)proton and (anti)neutron yields. Since $m_{n} \approx m_p$, and the neutron and proton both have spin-1/2, we have
\begin{align}
    \frac{\bar{p}}{p} &\approx \exp(-2(\mu_B + \mu_Q) / T) \\
    \frac{n}{p} &\approx \exp(-\mu_Q / T) \\
    \frac{\bar{n}}{p} &\approx \exp(-(2\mu_B + \mu_Q) / T)
\end{align}
So if we know $\mu_B / T$, $\mu_Q / T$, we can estimate the ratio of net-baryons to protons (not net-protons) as
\begin{align} \label{eq:bcorrection}
    \frac{B - \bar{B}}{p} &\approx \frac{p - \bar{p} + n - \bar{n}}{p} \\
    &\approx 1 - \exp(-2(\mu_B + \mu_Q) / T) + \exp(-\mu_Q / T) - \exp(-(2\mu_B + \mu_Q) / T)
\end{align}

The value of $\mu_B$ has been estimated in the literature, e.g. in Table VIII of  {Ref.} \cite{STAR:2017sal}, but $\mu_Q$ is not as well known. By examining deuteron yields, an estimate was provided in  {Ref.} \cite{STAR:2019sjh} for $\mu_Q / T$ as a function of $\sqrt{s_{\rm NN}}$. From Table \ref{tab:qn}, 
\begin{align} \label{eq:likeratio}
    \frac{\bar{d}}{d} &= \exp((-4\mu_B - 2\mu_Q) / T) \\
    \frac{\bar{p}}{p} &= \exp((-2\mu_B - 2\mu_Q) / T)
\end{align}
(the mass-dependent factors cancel) so
\begin{equation} \label{eq:qratio}
    \frac{\bar{d} / d}{\bar{p}^2 / p^2} = \exp(2\mu_Q / T)
\end{equation}
or as given in  {Ref.}  \cite{STAR:2019sjh},
\begin{equation} \label{eq:muQ}
    \frac{\mu_Q}{T} = \frac{1}{2} \ln \left( \frac{\bar{d} / \bar{p}^2}{d / p^2} \right)
\end{equation} 
As a caveat,  {Ref.} \cite{STAR:2019sjh} doesn't have data for 7.7 GeV (or 14.5 GeV), and also the deuteron yields (and thus the corresponding $\mu_Q / T$ values) have large errors.

\begin{figure}[!htbp]
    \centering
    \includegraphics[width=0.6\linewidth]{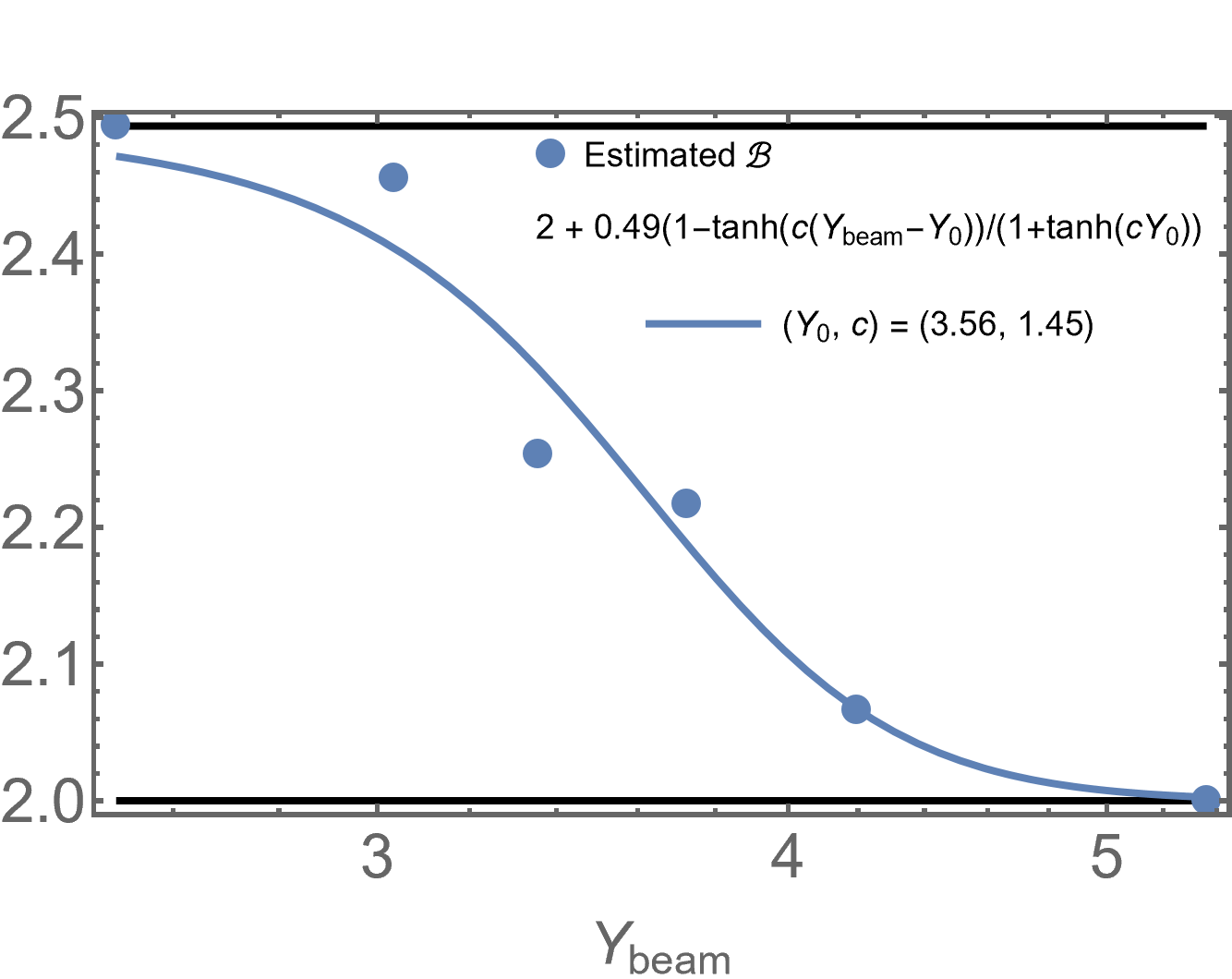}
    \caption{Estimated values of the baryon correction factor $\mathcal{B} \equiv (B - \bar{B}) / (p - \bar{p})$, together with the fit from Eq.~\eqref{eq:bcfinterp}. The black lines are the theoretical limits of 2 (isospin symmetry) and $197 / 79$ (from the initial baryon-to-proton ratio).}
    \label{fig:bcf}
\end{figure}

We calculate a correction factor to convert net-proton to net-baryon number as follows. There are two independent parameters, $\mu_B / T$ and $\mu_Q / T$, that vary with beam energy. For $\mu_B / T$, we use the estimates of $\mu_B$ and $T$ provided in Table VIII of  {Ref.} \cite{STAR:2017sal}, for 0-5\% centrality. For $\mu_Q / T$, we use \eqref{eq:muQ}, with (anti)deuteron data from  {Ref.} \cite{STAR:2019sjh}. Then, with both $\mu_B$ and $\mu_Q$, we can calculate the correction factor using \eqref{eq:bcorrection}. Note that the denominator in that equation is the number (or specifically rapidity density at mid-rapidity) of protons, not net-protons. Thus, we divide by the quantity $1 - \bar{p} / p$ to get the baryon correction factor
\begin{equation}
    \mathcal{B} \equiv \frac{B - \bar{B}}{p - \bar{p}}
\end{equation}
Finally, we'd expect $\mathcal{B}$ to stay bounded between 2 (which is the limit of perfect isospin symmetry) and $197 / 79$ (which is inherent from the initial Au nucleus). If $\mathcal{B}$, as calculated above, doesn't fall within this range, it is rounded to the nearest of the two values. 
The so-obtained baryon correction factor values, where experimental data are available, are shown as solid spherical symbols in Fig.~\ref{fig:bcf}.

Next, we can make a smooth interpolation of these results. Since the bounds to the baryon correction factor are assumed, we can choose a function with free parameters that interpolates between these bounds, fitting to the correction factors extracted from data. We choose the following form:
\begin{equation} \label{eq:bcfinterp}
    \mathcal{B} = 2 + 0.49 \cdot \frac{1 - \tanh(c(Y_{\rm beam} - Y_0))}{1 + \tanh(cY_0)}
\end{equation}
The ``center'' and ``slope'' parameters are $(Y_0, c) = (3.56, 1.45)$ respectively from the fitting analysis. In  {Fig.~}\ref{fig:bcf}, the interpolated curve is shown with its fit data. This interpolation result is used to compare model predictions with experimental measurements for calibrating the model parameters $\bar{e}$ and $\sigma$ in Sec.~\ref{subsec:calib}. 

\bibliography{ref-new.bib}

\end{document}